\documentclass[12pt,a4paper,twoside,groupcitations]{article}
\usepackage[T1]{fontenc}
\usepackage[ansinew]{inputenc}
\usepackage[english]{babel}
\usepackage{amsfonts,bm}
\usepackage{amsmath}
\usepackage{array}
\usepackage{amsthm}
\usepackage{amssymb}
\usepackage{graphicx}
\usepackage{subfigure}
\usepackage{braket}
\usepackage{eucal}
\usepackage{verbatim}
\usepackage[table]{xcolor}
\usepackage{caption}
\usepackage{cite}
\usepackage{textcomp}
\usepackage{multicol}
\usepackage{tikz}
\usetikzlibrary{positioning,arrows}
\usetikzlibrary{decorations.pathmorphing}
\usetikzlibrary{decorations.markings}
\usetikzlibrary{calc,decorations.markings}
\usetikzlibrary{arrows,shapes}
\usetikzlibrary{matrix,arrows}
\usepackage{pgfplots}
\usepackage{xparse}
\definecolor{jade}{HTML}{00A86B}
\newcommand{\be}{\begin{eqnarray}}
\newcommand{\ee}{\end{eqnarray}}

\renewcommand{\d}{\mbox{${\rm d}$}} 

\newcommand{\beq}{\begin{equation}}
\newcommand{\eeq}{\end{equation}}

\title{\bf Absence of covariant singularities in pure gravity}
\author{Roberto~Casadio$^{ab}$\thanks{E-mail: casadio@bo.infn.it},
$\ $
Alexander~Kamenshchik$^{abc}$\thanks{E-Mail: kamenshchik@bo.infn.it},
$\ $
and
Iber\^e Kuntz$^{ab}$\thanks{E-mail: kuntz@bo.infn.it}
\\
\\
$^a${\em Dipartimento di Fisica e Astronomia, Universit\`a di Bologna}
\\
{\em via Irnerio~46, 40126 Bologna, Italy}
\\
\\
$^b${\em I.N.F.N., Sezione di Bologna, I.S.~FLAG}
\\
{\em viale B.~Pichat~6/2, 40127 Bologna, Italy}
\\
\\
$^c${\em L.D. Landau Institute for Theoretical Physics}
\\
{\em 119334 Moscow, Russia}
}
\begin{document}
\date{}
\maketitle
\begin{abstract}
The assumptions of the Hawking-Penrose singularity theorem are not covariant under field redefinitions.
Following the works on the covariant formulation of quantum field theory initiated by Vilkovisky and DeWitt
in the 80's, we propose to study singularities in field space, where the spacetime metric
is treated as a coordinate along with the other fields in the theory.
From this viewpoint, a spacetime singularity might be just a singularity in the field-space coordinates,
analogously to the standard coordinate singularities in General Relativity.
Objects invariant under field-space coordinate transformations can then reveal whether certain spacetime
singularity is indeed singular. We recall that observables in quantum field theory are scalar functionals
in field space.
Therefore, in principle, spacetime singularities corresponding to regular field-space curvature invariants
would not affect physical observables.
In this paper, we show that the field-space Kretschmann scalar for a certain choice of the
DeWitt~field-space metric is everywhere finite.
This fact could be interpreted as an indication that  no singularities actually exist in pure gravity
for any gravitational action.
In particular, all vacuum singularities of General Relativity result from an unhappy  choice of field variables.
The extension to the case in which matter fields are present, as required by singularity theorems,
is left for future development.
\par
\null
\par
\noindent
\end{abstract}
\newpage
\section{Introduction}
\setcounter{equation}{0}
\label{Sintro}
Spacetime singularities are a trademark of General Relativity~\cite{HE}.
In static and stationary spacetimes, these singular regions must be excised from the classical spacetime
manifold~\cite{geroch}, thus affecting the spacetime topology. 
What makes them more interesting is their possible occurrence at the end of the gravitational collapse or at the beginning
of the cosmological expansion, when the energy density of matter would classically grow unbounded.
In fact, mathematical theorems predict the general occurrence of geodesic incompleteness in the
classical theory of gravity when matter satisfies suitable energy conditions~\cite{HE}. 
However, there are indications that this unlimited growth does not survive the quantization of any toy model
which is simple enough to be analysed exactly~\cite{OS}, or even approximately~\cite{WKB}.
As the Heisenberg uncertainty principle naively suggests, the gravitational pull is overbalanced by the quantum
pressure which prevents matter to be localised too narrowly, albeit the scale of this ``bounce'' remains hard to
estimate reliably.
\par
Another important result applying to high energy density scenarios is obtained by treating the Einstein-Hilbert action
as a field theory on a fixed background.
In this instance, loop corrections switch on new interaction terms, which can be viewed as a quantum completion
of the initial theory~\cite{Parker:2009uva}.
Conversely and more appropriately, one can regard the Einstein-Hilbert action as the effective low-energy theory
which holds when the extra terms are small with respect to the Planck scale and study loop effects on the singularity
theorems~\cite{Kuntz:2019lzq,Calmet:2020vuh,Kuipers:2019qby}.
\par
At the same time, attempts at ``reconciling'' the presence of singularities 
with the classical theory of General Relativity have also been undertaken.
In particular, by reconciliation we mean that points corresponding to singularities are not excluded
from the spacetime manifold but, instead, mechanisms for crossing these points are elaborated.
With regard to such mechanisms, it is useful to recall that, besides ``strong'' singularities, such as
the Big Bang--Big Crunch in cosmology or the Schwarzschild central singularity,
there exist also the so called ``soft'' or sudden
singularities~\cite{soft,soft1,soft2,soft3,Fern,Fern1,paradox1,my-review,Kamen-Specola}.
The particularity of soft singularities is that, while curvature invariants diverge,
the Christoffel symbols remain finite (or even vanish, like for the Big Brake~\cite{soft3}),
hence geodesics are well defined and can be extended through the singularity itself.
A complete spacetime is so reconstructed that the singularity crossing takes place in it.
The lesson that one can extract from these cases is that the presence of curvature singularities 
does not always represent too grave of a menace.
\par
The idea of resolving or passing through a strong singularity
of the Big Bang--Big Crunch type or of the Schwarzschild type, looks much more counterintuitive
with respect to the crossing of soft singularities.
Nevertheless, a number of works devoted to this topic have been published
recently~\cite{Wetterich:2013aca,Wetterich:2014zta,Wetterich:2020oyy,Naruko:2019gsi,Domenech:2019syf,
Kam16,Kamenshchik:2017ojc,Kamenshchik:2018crp,Bars,Bars1,Kozl,Mercati,Sloan,Mercati:2021zmv,Prester,Prester1}. 
From our point of view, it seems that all of these approaches resort to one of two ideas, or a combination
thereof.
One of these ideas is to employ a reparameterization of the field variables which makes
the singular geometrical invariant non-singular.
Another idea is to find such a parameterization of the fields, including, naturally, the metric,
that gives enough information to describe consistently the crossing of the singularity even
if some of the curvature invariants diverge.
It must be said that there is no consensus about approaches describing the crossing of
singularities of the Big Bang--Big Crunch type, and sometimes the debate escalates.
In Refs.~\cite{Bars,Bars1} the procedure for the crossing of the Big Bang--Big Crunch
singularity based on the use of Weyl symmetry, was elaborated. 
Using a Weyl-invariant theory, with two scalar fields conformally coupled to gravity,
the authors obtained the geodesic completeness of the corresponding spacetime.
The consequence of this geodesic completeness is the crossing of the Big Bang
singularity and the emergence of antigravity regions in the Einstein frame. 
The use of Weyl symmetry to describe the passage through the Big Crunch--Big Bang
singularity accompanied by a change of sign for the effective Newton constant,
has led to some discussion.
In Ref.~\cite{polem,polem1}, it was noticed that for such a passage through the singularity
some of the curvature invariants still become infinite.
In Ref.~\cite{Bars1} a counter-argument was put forward according to which,
if one has enough conditions so as to match the nonsingular quantities before and
after crossing the singularities, then the singularities can be traversed.
\par
Given the complexity of the singularity problem in General Relativity,
and in modifications or generalisations thereof, we think that it makes sense to try and
take the complementary point of view of analysing the structure of singularities
in the functional space of all field configurations.
This is indeed lined up with the covariant approach to quantum
field theory~\footnote{That physics should remain the same under field
reparameterizations has also been called ``field relativity'' by Wetterich~\cite{Wetterich:2013aca}.
This idea indeed goes back to the works of Vilkovisky and DeWitt in the early 80's,
where they introduced a metric and a connection in field space to enforce reparameterization
invariance at the quantum level.
Wetterich states ``field relativity'' as a principle, without showing how one can obtain an invariant
effective action.
We also take it as a principle, but within Vilkovisky-DeWitt's formalism for the effective action.
That being said, our work here concerns the classical regime, thus one needs not dwell on this more
complicated subject.}~\cite{Vilkovisky:1984st,DeWitt:1988dq,DeWitt:2003pm},
which builds up on the field-space formulation of canonical quantum gravity as proposed in DeWitt's
seminal
paper~\cite{DeWitt:1967yk}. We shall, however, work in the classical limit and leave the generalization
to the quantum regime for another work.
It is particularly tempting to investigate whether the absence
of spacetime singularities could be a property of the space of fields of the complete theory
of quantum gravity.
This requires understanding the connection between spacetime singularities and singular
points in field space.
\par
Field-space singularities are linked to spacetime singularities once a choice for the parameterisation of fields
is made.
This choice, however, is not unique.
One can in fact perform field redefinitions, as customary in field theory, without affecting the physical observables
of the theory.
The freedom of choosing the field parameterisation allows for the removal of spacetime singularities in some cases.
This naturally leads to distinct types of singularities, classified in accordance with the possibility of removing
them via field redefinitions.
Our perspective is that removable spacetime singularities under field redefinitions reflect an initial bad
choice of field parameterisation, like removable spacetime singularities reflect a bad choice of coordinates.
\par
It is not always easy, however, to find a convenient  field redefinition (if it exists) able to remove spacetime
singularities
 without introducing other defects.
Clearly, one cannot conclude that such singularities are non-removable before exhausting all infinite possibilities
of field parameterisation.
A covariant approach under field redefinitions is thus of utmost importance and a practical necessity.
In this paper, we shall investigate singularities in field space by adopting a field-covariant approach.
The main result of this paper is the calculation of the field-space Kretschmann scalar in pure gravity,
which turns out to be constant and free of singularities.
\par
This paper is organised as follows.
In Section~\ref{Ssuper}, we review coordinate singularities in spacetime and use the same reasoning
to classify singularities in field space.
In Section~\ref{Sexample}, we review one known example, namely the Hawking-Turok instanton, where singularities can be removed
by field redefinitions.
These removable field-space singularities motivate the search of a more general formalism,
which we start to develop in the rest of the paper.
We show in Section~\ref{Ssingularity} that the assumptions in the Hawking-Penrose theorem
are not invariant under field redefinitions, urging the investigation of singularities in the field space.
Section~\ref{Sgeometry} is devoted to the geometry of field space for pure gravity, which is then used to
study singularities in a field-covariant way.
We discuss our results in Section~\ref{Sconc}.
\par
Before continuing, we must remark a potential source of confusion due to the differences in nomenclature
adopted in the literature when it comes to the space of all fields.
Field space, superspace, space of histories, configuration space and functional space usually refer to some
notion of ``set of all fields''.
Sometimes these terms are used interchangeably and other times they are used with different meanings,
{\em e.g.}~superspace might refer to the space of three-dimensional fields in the ADM formalism~\cite{ADM}
or the space of fields defined on the entire four-dimensional spacetime.
To avoid misunderstandings, throughout this paper we shall adopt the term ``field space'' to denote the set of
all possible field configurations in the entire spacetime manifold.
This set is infinite dimensional and is not restricted to solutions of the classical equations
of motion.~\footnote{As we shall see below, field-space singularities will correspond to points
in field space that are solutions of some equations of motion. The formalism is nonetheless general
and independent of the particular theory that one has in mind.}
The adjectives ``functional'' and ``field-space'', as in ``functional tensor'' or ``field-space tensor'',
shall both be used interchangeably to refer to objects or concepts in the field space.
\section{Field space and singularities}
\setcounter{equation}{0}
\label{Ssuper}
Like the principle of General Relativity says that physics does not depend on the spacetime coordinates,
it is widely accepted that observables (like cross sections) should not depend on the fields parametrisations
we use to describe physical processes~\footnote{Expectation values, and more general correlation
functions, are also observables in a quantum theory (at least formally).
However, we should again stress that we focus on the classical level here, for which
examples of observables include the field-space curvature invariants, such as the field-space Kretschmann scalar
(see Section~\ref{Sgeometry}).
These objects are rather abstract and it is difficult to conceive a way to measure them directly.
One thus needs to relate known quantities measured in a laboratory to field-space curvature invariants.
At the quantum level, on the other hand, the Vilkovisky-DeWitt formalism guarantees that all correlation
functions remain covariant under field redefinitions.
\label{f3}}. 
The field space $\mathcal S$ was developed in order to treat on the same (geometrical) footing both changes of coordinates
in the spacetime $\mathcal M$ and field redefinitions in the functional approach to quantum field theory
(for a review, see {\em e.g.}~Refs.~\cite{Parker:2009uva,DeWitt:2007mi}).
In fact, the action can be written as a functional on $\mathcal S$ in a way that makes it explicitly invariant
under spacetime diffeomorphisms and small field variations.~\footnote{Note the technical and substantial fact that
coordinate transformations and field redefinitions are not (and should not - more later) be restricted
to the ``small'' ones which can be connected with the identity when singularities are present.}
For this purpose, all field components are denoted by indices which represent both the
spacetime point and the field and tensor components (the so-called DeWitt condensed
notation~\cite{DeWitt:2003pm}).
{We shall thus reserve capital Latin letters for discrete indices ({\em e.g.}, tensorial, gauge and spinor indices),
Greek letters shall correspond to usual spacetime indices and lower-case Latin letters will denote the union of all indices
(discrete and continuum, the latter being denoted by spacetime coordinates such as $x^\mu$ or $y^\mu$).}
For example, if the theory contains a scalar field $\phi$ and the metric $\bm g$, one collectively writes
both fields together as
\be
\varphi^a
=
\varphi^{(A,\mu)}
=
\varphi^A(x^\mu)
=
\left(\phi(x^\mu), g^{\alpha\beta}(x^\mu)\right)
\ .
\ee
Note that field reparameterizations can be viewed as a particular case of field-space transformations
acting on the index $A$ (leaving coordinates unaffected), whereas changes in the spacetime
coordinates act both on the argument $x^\mu$ and the tensor indices included in $A$.
\par
All results regarding coordinate invariance in a given spacetime manifold and field parameterisation
should be automatically accounted for in this formalism. 
However, this is partly hindered by the convention which assumes spacetime integration
for repeated indices.
For example,
\be
\varphi^a\,\varphi_a
=
\sum_{B\in \mathcal A}
\int_{B}
\varphi^A(x)\,\varphi_A(x)\,\d\mu(x)
\ ,
\label{trace}
\ee
where $(B,x^\mu)$ are suitable charts of the atlas $\mathcal A$ for the spacetime $\mathcal M$ and
$\d\mu$ a suitable spacetime measure.
Including the above integration is convenient for the purpose of writing functionals in $\mathcal S$,
but makes it more problematic to analyse the nature of local singularities in spacetime. 
Moreover, one typically has $\d\mu=\sqrt{-g}\,\d^4 x$, where $g={\rm det}(\bm g)$ so that the
metric $\bm g$ would always play a special role (even if it were not dynamical).
\par
Finally, it is important to remark that the topology of $\mathcal M$ will affect the possible
metrics that belong to the gravitational sector of the field space
\be
\mathcal G(\mathcal M)
=
\frac{{\rm Lor}(\mathcal M)}{{\rm Diff}(\mathcal M)}
\ ,
\ee
where ${\rm Lor}(\mathcal M)$ denotes Lorentzian metrics and ${\rm Diff}(\mathcal M)$ the
diffeomorphisms on $\mathcal M$.
The existence of singularities of the kind predicted by the singularity theorems should therefore 
reflect in the topology of $\mathcal M$ and, consequently, the global structure of 
$\mathcal S\supseteq\mathcal G$. 
Moreover, the field content of the theory is also an ingredient that can be set independently but which
will affect the local and global features of both $\mathcal M$ and $\mathcal S$.
In the following we shall therefore consider the topology of $\mathcal M$ and the specific fields
as parts of what needs to be determined compatibly with local scalar quantities in $\mathcal S$.
\subsection{Spacetime singularities: known facts}
We know pretty much everything about this kind of singularities from textbooks in General Relativity,
where singularities are analysed for specific solutions of field equations.
The general definitions can however be given without assuming the metric tensor and other fields
satisfy equations of motion.
In particular, we can consider the following two types of singularities~\cite{HE}.
\subsubsection{Removable (coordinate) singularities}
Given a spacetime chart $(B,x^\mu)$, there might be field components (of the metric or otherwise)
which behave badly (diverge or vanish) on the border $\partial B$, although all local spacetime
scalar quantities remain smooth there. 
In this case, we expect that the singularities in field components can be removed by a suitable change 
of coordinates $x^\mu\to y^\mu(x^\nu)$ about the border of $B$, so that the new chart
$(\bar B,y^\mu)$ covers $\partial B$ and extends past the singularities.
\par
Note that the fact a singularity is removable does not necessarily mean it is physically irrelevant.
Typical example of a purely accidental bad behaviour is given by the origin $r=0$ in spherical coordinates,
where the angular components of the flat Euclidean metric vanish.
This singularity can of course be removed by changing to Cartesian coordinates, which cover the point $r=0$.
A different example is given by the Schwarzschild radius in the usual Schwarzschild coordinates,
where the metric $\bm g$ reads
\beq
\d s^2 =
 -\left(1-\frac{2\,M}{r}\right) \d t^2
 + \left(1-\frac{2\,M}{r}\right)^{-1} \d r^2 
 + r^2\,\d\Omega^2
\ .
\label{schw}
\eeq
The time component of the above covariant metric vanishes and the radial component diverges
for $r\to 2\,M$.
This singularity can also be removed by a change of coordinates, but its deep physical meaning
as an event horizon can only be assessed by studying the behaviour of geodesics~\cite{HE}.
After all, the role of the metric is precisely to define geodesic motion, that is to describe
gravity according to the equivalence principle.
Finally, note that any change of coordinates which removes the singular behaviour in the 
components of the metric~\eqref{schw} around $r=2\,M$ must depend on the mass $M$,
\be
x^\mu
\to
y^\mu(x^\nu,M)
\ ,
\ee
and cannot be a diffeomorphism, in that there exists no parameter which smoothly connects the above change
of coordinates to the identity.~\footnote{One should not consider $M\to 0$ as the way
to connect to the identity, since changing $M$ means to change the spacetime metric.} 
(One usually refers to such changes of coordinates as ``large''.)
This example is a nice reminder that the symmetry group of General Relativity (meaning general coordinate
invariance, not the dynamics of the particular Einstein-Hilbert action) is the Bergmann-Komar group~\cite{BK}
\be
x^\mu
\to
y^\mu(x^\nu,\bm g)
\ee
which includes, but is not exhausted by, spacetime diffeomorphisms.
\subsubsection{Non-removable singularities}
Changes of coordinates, either small or large, cannot affect spacetime scalar quantities.
Therefore, if there exists scalar quantities which behave badly at the border $\partial B$ of the spacetime
chart $(B,x^\mu)$, such singularities cannot be eliminated by defining a new chart $(\bar B,y^\mu)$ and one
must conclude that $\partial B\subseteq\partial\mathcal M$.
We remark once more that local (scalar) quantities are used to probe and determine the global structure
of $\mathcal M$.
\par
The typical examples are given by the origin $r=0$ in the Schwarzschild spacetime, where the Kretschmann
scalar diverges, and the big bang singularity in cosmology, where the scale factor of the Friedmann-Lema\^itre-Robertson-Walker (FLRW) metric vanishes.
In both cases, those singular points do not belong to $\mathcal M$, which is therefore said to be
geodesically incomplete.
These non-removable singularities are the subject of Penrose-Hawking theorems which prove their
general existence in General Relativity, albeit under suitable auxiliary conditions about the matter fields.
Note that curvature singularities cannot generally be seen as a consequence of the
Hawking-Penrose theorem.
In fact, nowhere in the theorem's proof appears divergent energy densities.
On physical grounds, however, it is reasonable to conjecture that both concepts are somehow related.
In practice, one can adopt such a conjecture by puncturing the points in spacetime where curvature
invariants diverge.
When this is performed, geodesics terminate at such missing points for a finite value of the affine
parameter/proper time, thus the spacetime becomes geodetically incomplete.
In this sense, geodesic incompleteness would capture both divergent curvatures and conical singularities
(in which geodesic incompleteness is present but the curvature remains regular).
In this paper, we shall adopt this point of view.
\subsection{Field-space singularities: a conjecture}
Since the field-space formalism includes the spacetime tensor structure, the above classification 
remains valid, although somewhat hindered by the convention~\eqref{trace} if the singularity is
localised (like in the Schwarzschild manifold).~\footnote{Cosmology is again simpler because one
assumes very strong symmetries like homogeneity, which makes the spacetime integration
somewhat less harmful.}
There is however hope that some of the spacetime singularities which cannot be removed by 
changes of coordinates could still be removed by field redefinitions.
Note that removing such singularities, strictly speaking, changes the topology of spacetime and
is therefore tantamount to extend the singular spacetime manifold $\mathcal M_1$ to a
different $\mathcal M_2$ which includes the originally singular points.
It now seems natural to consider local quantities behaving like scalars under field reparameterisations
to likewise define removable and non-removable field-space singularities.
\par
This approach requires introducing a local metric $\bm G$ in field space $\mathcal S$ and computing
the associated geometric scalars by defining a covariant derivative which is compatible with $\bm G$.
Note that $\bm G$ is actually determined by the kinetic part of the action and its dimension depends
on the field content of the latter.
There is therefore a huge freedom in defining $\bm G$ and one might hope that adding enough
fields always allows one to remove all divergences.
However, classification of such high dimension metric manifolds with respect to singularities
is not complete and only specific examples can be worked out.
\par
A few more remarks are also in order.
Firstly, like for spacetime singularities, it is clear that ``large'' field redefinitions will be needed to
remove singularities.
Moreover, it is also clear that such field redefinitions will have to depend on the specific configuration
we wish to cure, so that the relevant transformations generalise the Bergmann-Komar ones to include
a dependence on all fields in the specific configuration.
Secondly, note that none of the above assumes or needs the field configurations to satisfy any equations
of motion.
The fact that $\bf G$ is related to the kinetic term in the action simply reflects the requirement that
a canonical choice for the fields should exist locally.
\par
\section{Hawking-Turok instanton}
\setcounter{equation}{0}
\label{Sexample}
Several examples in the literature suggest that some non-removable spacetime singularities
can still be removed by field redefinitions~\cite{Wetterich:2013aca,Wetterich:2014zta,Naruko:2019gsi,
Domenech:2019syf,Kamen-Specola,Kam16,Kamenshchik:2017ojc,Kamenshchik:2018crp,Bars,Bars1}.
In this section, we briefly review one such example in cosmology, namely the Hawking-Turok
instanton describing an open universe obtained by analytical continuation of a singular
Euclidean spacetime~\cite{Hawking:1998bn}.
Hawking and Turok have pointed out that a singular solution might be acceptable should the on-shell
action evaluated at such a solution be regular~\cite{Hawking:1998bn}.
At the level of classical General Relativity, this implies that only the spacetime Ricci scalar and matter fields
are required to be finite.
This is indeed the case for the aforementioned instanton, as we shall now see.
\par
The Euclidean action in this model reads
\beq
S_E
=
\int\mathrm{d}^4x\, \sqrt{g}
\left[
-\frac12 \,R + \frac12\, \nabla_\mu\phi\,\nabla^\mu\phi + V(\phi)
\right]
\ ,
\label{eq:ht}
\eeq
where $R$ is the Ricci scalar for the metric $\bm{g}$, $\phi$ is a canonical scalar field and $V(\phi)$
its potential.
The equations of motion corresponding to Eq.~\eqref{eq:ht} contains the solution
$\phi = \phi(\sigma)$ and
\beq
\d s^2
=
\d\sigma^2 + b^2(\sigma) \left[\d\chi^2 + \sin^2(\chi)\,\d\Omega^2\right]
\ ,
\label{eq:solht}
\eeq
where the factor
\beq
b(\sigma) \approx
\begin{cases}
	\sigma\ ,
	& \text{for} \quad \sigma \sim 0
	\ ,
	\\
	(\sigma_f - \sigma)^{1/3}\ ,
	& \text{for} \quad \sigma \sim \sigma_f
	\ ,
\end{cases}
\eeq
and the scalar field
\beq
\phi(\sigma) \approx
\begin{cases}
	\frac12\,{\sigma^2}
	\ ,
	& \text{for} \quad \sigma \sim 0
	\ ,
	\\
	-\sqrt{\frac23} \ln(\sigma_f - \sigma), 
	& \text{for} \quad \sigma \sim \sigma_f
	\ .
\end{cases}
\eeq
After Wick rotating $\chi$, one then obtains an open universe.
The Ricci scalar for the solution~\eqref{eq:solht} reads
\beq
R \sim \frac{1}{(\sigma_f - \sigma)^2}
\ ,
\label{eq:cosmosing}
\eeq
thus there is a non-removable spacetime singularity at $\sigma = \sigma_f$.
The on-shell action is nonetheless regular should the potential $V(\phi)$ not grow faster than $b^3$
for $\sigma \to \sigma_f$:
\beq
S_E|_\text{on-shell}
=
-\int\mathrm{d}^4x \, b^3(\sigma) \, V(\phi)
\ ,
\eeq
where $S_E|_\text{on-shell}$ denotes the action~\eqref{eq:ht} evaluated at the solution~\eqref{eq:solht}.
Because the action is a scalar functional in the field space, the finiteness of $S_E|_\text{on-shell}$
suggests that the singularity observed in Eq.~\eqref{eq:cosmosing} should instead be removable
by field redefinitions.
In fact, a simple rescaling of the metric is able to remove this singularity as we now explain.
\par
Changing spacetime coordinates to $\d\bar\sigma = b^{-1} \,\d\sigma$, followed by a Weyl transformation
$\bar g_{\mu\nu} = b^{-2}\, g_{\mu\nu}$, leads to
\beq
\d s^2
=
b^2(\bar\sigma)\, \d\bar{s}^2
\ ,
\eeq
where we defined
\beq
\d\bar{s}^2
=
\d\bar\sigma^2 + \d\chi^2 + \sin^2(\chi)\,\d\Omega^2
\ .
\label{eq:flat}
\eeq
Near the singularity, that is for $\sigma \sim \sigma_f$, one has
\beq
\bar\sigma
=
\frac32 \left(\sigma_f - \sigma\right)^{2/3}
\ ,
\eeq
and the new geometry $\bar {\bm{g}}$ is flat and obviously regular.
However, the scalar field 
\beq
\phi = -\sqrt{6}\,\log b(\bar{\sigma})
\eeq
becomes singular as $b(\bar{\sigma}) \to 0$ (for $\bar{\sigma} \to 0$).
We have thus shifted the singularity from the geometry to the scalar field.
\par
In order to remove the singularity from the scalar field as well, one can perform another
Weyl transformation~{\cite{Domenech:2019syf}}
\beq
\bar g_{\mu\nu}
=
\Omega^2\, \tilde g_{\mu\nu}
\ ,
\eeq
with 
\beq
\Omega = 1 + \beta\, e^{-\alpha \,\sqrt{2/3}\, \phi}
\ .
\eeq
where $\alpha$ and $\beta$ are free parameters.
Note that the absence of singularity in the geometry $\bar{\bm{g}}$
is not spoiled by this additional transformation as it goes to unity at the singularity ($\phi \to \infty$).
Near the singularity at $\bar{\sigma} = 0$, the new Ricci scalar takes the form
\beq
\tilde R
\sim
\bar{\sigma}^{\alpha - 2}
\ , 
\eeq
thus it is regular for $\alpha > 2$.
The action in the field frame $\tilde{\bm g}$ becomes
\begin{eqnarray}
S_E
&\!\!=\!\!&
\int\d^4x\, \sqrt{\tilde g}
\left[
 -\frac12 \,e^{-\sqrt{2/3}\,\phi}\, \Omega^2\, \tilde R
+ 3\, e^{-\sqrt{2/3}\,\phi}\, \Omega^2\, 
\frac{\partial\ln\Omega}{\partial \phi}
\left(\sqrt{\frac{2}{3}} - \frac{\partial\ln\Omega}{\partial\phi}\right)
\tilde \nabla_{\mu} \phi\,\tilde \nabla^\mu \phi
\right.
\nonumber
\\
&&
\phantom{\int\d^4x \sqrt{\tilde g}}
\left.\phantom{\frac AB}
+ \Omega^4 \,\bar V(\phi)
\right]
\ .
\end{eqnarray}
Now we can see that the canonically-normalised scalar field
\beq
\d\tilde\phi^2
=
6\, e^{-\sqrt{2/3}\,\phi}\,\Omega^2\,
\frac{\partial\ln\Omega}{\partial\phi}
\left(\sqrt{\frac{2}{3}} - \frac{\partial\ln\Omega}{\partial \phi}\right)
\d\phi^2
\eeq
is also regular in the spacetime with metric $\tilde{\bm  g}$.
In fact, near the singularity at $\bar{\sigma} = 0$, one finds
\beq
\tilde\phi
\sim
\bar{\sigma}^{(1+\alpha)/2}
\ ,
\eeq
where we set $\beta = -1/4\alpha$.
Therefore, in conformity with the finitude of $S_E|_\text{on-shell}$, we have found a field-space frame
where both the geometry and the scalar field are singularity-free.
More precisely, the above sequence of field transformations for the metric and scalar field allows one
to extend the spacetime manifold to include the singularity $\sigma = \sigma_f$,
which is therefore removable in field space.
\par
It is easy to see that the sequence of field transformations mapping $\varphi^a=(\phi,\bm g)$ into
$\tilde\varphi^a=(\tilde\phi,\tilde{\bm g})$
does not map all spacetime geodesics of the original metric $\bm g$ in Eq.~\eqref{eq:solht} into geodesics 
(straight lines) of the flat metric $\tilde{\bm g}$ in Eq.~\eqref{eq:flat}.
Therefore, one runs into the problem of having to decide {\em a priori\/} whether test particles
{({\em e.g.}, the classical and non-relativistic electron)} fall freely in the metric $\bm g$
or in the metric $\tilde{\bm g}$.
This ambiguity is removed {in a fundamental theory where test particles are fully replaced
by matter fields (such as spinors in the example of the electron), in which case the matter content of
the Hawking-Turok model is solely given by the scalar field $\phi$.}
In that case, the entire dynamics is unique and one does not need to study spacetime geodesics at all,
although the distinction between matter and gravity is lost under the field redefinitions.
The notable exception is given by trajectories of constant $\chi$, which
are geodesics in both field frames.
However, the expansion of congruences of such geodesic, namely
\be
\theta
=
u^\mu_{\ ;\mu}
\ ,
\label{eq:exp}
\ee
clearly vanishes in $\tilde{\bm g}$ and takes the well-known FLRW form~\cite{Raychaudhuri:1953yv}
\be
\theta
=
3\,\frac{b'(\sigma)}{b(\sigma)}
\ee
in the initial metric $\bm g$.
There is no geodesic focusing in $\bm g$, whereas the Hawking-Penrose theorem applies in $\tilde{\bm g}$.
This exemplifies the issue of field frame dependence of the singularity theorems we are going to discuss next.
\section{Functional frame dependence of Hawking-Penrose theorem}
\setcounter{equation}{0}
\label{Ssingularity}
The causal structure of a given spacetime can be inspected by studying congruences of geodesics,
whose behaviour is determined by the Raychaudhuri equation~\cite{Raychaudhuri:1953yv}.
In particular, the positivity of the discriminant
\beq
\Delta
=
R_{\mu\nu}\,u^\mu \,u^\nu
\ ,
\eeq
where $\bm u$ is the geodesic four-velocity field and $\bm R$ the Ricci tensor, ensures the focusing property of gravity,
that is, the fact that the expansion~\eqref{eq:exp} of time- and light-like congruences diverges at focal points.
The Hawking-Penrose theorem~\cite{HE} then relies on the positivity of $\Delta$ (along with other assumptions)
for light-like geodesics in order to further establish the geodesic-incompleteness of spacetime.
Nonetheless, $\Delta$ is not covariant under field redefinitions and the focusing property can therefore be affected
by changing the field frame.
This fact can already be seen by considering a pure metric redefinition $\bm{g}\to\tilde{\bm{g}}$ 
(of course, not corresponding to a change of spacetime coordinates, but not involving other fields either),
under which the Ricci tensor transforms as
\beq
\tilde R_{\mu\nu}
=
R_{\mu\nu} + A_{\mu\nu}
\ ,
\eeq
with
\beq
A_{\mu\nu} 
=
2\, C^\sigma_{\ \nu[\mu} \,C^\rho_{\ \rho]\sigma}
- 2\,\nabla_{[\mu} C^\rho_{\ \rho]\nu}
\ ,
\eeq
where
\beq
C^\alpha_{\ \mu\nu}
=
\frac12\, \tilde g^{\alpha\beta}
\left(\nabla_\mu\tilde g_{\beta\nu} + \nabla_\nu \tilde g_{\beta\mu} - \nabla_\beta \tilde g_{\mu\nu}\right)
\ ,
\eeq
and covariant derivatives are taken with respect to the original metric $\bm g$.
It is precisely $A_{\mu\nu}$ that makes the Ricci tensor a non-tensorial quantity under these restricted
field redefinitions.~\footnote{Note, however, that $A_{\mu\nu}$ is itself a tensor in field space.}.
\par
The focusing condition $\Delta> 0$, as much as the singularity theorems in their current form,
are thus not covariant under field redefinitions.
Therefore, the physical meaning of singularities in field space should be put on the same footing as 
(removable or non-removable) singularities of spacetime. 
In fact, we should also recall that ways of evading those theorems are known (see, {\em e.g.}~Ref.~\cite{bargueno}).
Strictly speaking, singularities implied by the Hawking-Penrose theorem are most likely field-space coordinate singularities,
unless one also has $A_{\mu\nu}\,u^\mu\,u^\nu>0$ or
\beq
2\, C^\sigma_{\ \nu[\mu} \,C^\rho_{\ \rho]\sigma}\,u^\mu\, u^\nu
>
2\,\nabla_{[\mu}\, C^\rho_{\ \rho]\nu}\,u^\mu\, u^\nu 
\ .
\eeq
This condition will depend on the metric redefinition one makes and it seems rather unlikely that it will hold
for any frame in field space without being too restrictive.
Resorting to energy conditions will not change this scenario because energy conditions also depend on the frame
in field space~\cite{Chatterjee:2012zh,Capozziello:2013vna}.
\par
We should also point out that the proper time
\beq
\Delta\tau
=
\int\sqrt{g_{\mu\nu}\,\d x^\mu\, \d x^\nu}
\eeq
is obviously dependent on the functional frame, thus the rate of change of the expansion $\theta$
for time-like geodesics has yet another instance of dependence on the field-space frame, and
so consequently do the focusing theorems in this case. 
Affine parameters generally depend on the field-space frame due to their definition via the geodesic
equation, which contains an implicit dependence on the metric.
Finally, the very definition of the expansion $\theta$, given in Eq.~\eqref{eq:exp} for any tangent vector $u^\mu$,
also spoils the covariance under field redefinitions because of the metric dependence in the
covariant derivative.
Therefore, there are three different instances of functional non-covariance in the Hawking-Penrose
theorems~\footnote{Similar conclusions can be found in Ref.~\cite{Wetterich:2014gaa}.}
originating from (1)~affine parameters, (2)~the geodesic expansion and (3)~the discriminant $\Delta$.
\par
Note that, at first sight, it is somewhat expected that the singularity theorem would not hold
under field redefinitions, as Einstein's equations certainly take a different form in different field-space
coordinates.
It is important to remark, however, that there are other sources of non-covariance in addition
to the equations of motion, as we outlined above.
Although the form of the equations of motion change, it could be the case that all sources of
non-covariance could cancel each other out, leaving the Hawking-Penrose theorem invariant.
This turns out not to be the case.
\par
Although the Hawking-Penrose theorem sets the conditions for the formation of spacetime singularities,
the above analysis shows that it cannot discriminate between non-removable and field-coordinate singularities.
We thus need an object that is invariant under both coordinate transformations and field redefinitions.
This invariably requires the study of the geometry of field space. 
\section{Geometry of field space for pure gravity} 
\setcounter{equation}{0}
\label{Sgeometry}
The results of the preceding Sections~\ref{Sexample} and~\ref{Ssingularity} suggest that fully non-removable
singularities, in addition to being coordinate-independent, must be independent of the choice we make to
describe the dynamical fields.
In this Section, we propose to look at local invariants in the field space $\mathcal S=\mathcal S(\mathcal M)$,
such as the functional Kretschmann scalar, to decide whether singularities are removable by field redefinitions.
\par
We shall now revert to the DeWitt condensed notation briefly recalled in Section~\ref{Ssuper},
with the field space parameterized by $\varphi^a$, which collectively denotes all fields of arbitrary spin
present in the theory.
These fields are supposed to be interpreted as mere coordinates in the field space $\mathcal S$.
Notice that the field space, as the set of all field configurations, comprises points that are solutions
and points that are not solutions of some field equations.
We nonetheless define a singularity $\varphi^a=\varphi^a_{\rm s}$ in field space as a solution of some field equations
for which the field-space curvature invariants are infinite at $\varphi^a_{\rm s}$.
Without specifying the theory, one can analyse the existence of possible field-space singularities by calculating
the curvature invariants with some choice of the field-space metric.
\par
We should also stress that under field redefinitions $\varphi^a \to \tilde \varphi^a$ the {classical} equations
of motion and their solutions transform in such a way that the redefined solutions will be solutions
of the redefined equations of motion.
This indeed reflects the fact that the {classical} action transforms as a scalar under field redefinitions
$\tilde S[\tilde \varphi^a] = S[\varphi^a]$, thus the {classical} equations of motion transform
covariantly,~\footnote{As we mentioned in footnote~\ref{f3}, this does not automatically hold
at the quantum level, since the standard effective action acquires off-shell corrections under field redefinitions.
The Vilkovisky-DeWitt effective action is precisely designed in order to keep the covariance of the effective
equations of motion.}
\beq
\frac{\delta \tilde S[\tilde\varphi]}{\delta\tilde\varphi^a}
=
\frac{\delta\varphi^b}{\delta\tilde\varphi^a} \,\frac{\delta S[\phi]}{\delta\varphi^b}
\ .
\eeq
This implies
\beq
\frac{\delta \tilde S[\tilde\varphi]}{\delta\tilde\varphi^a}
=
0
\quad
\Leftrightarrow
\quad
\frac{\delta S[\varphi]}{\delta\varphi^b}
=
0
\ ,
\eeq
which means that solutions in one field-space coordinates will be taken into solutions in
another field-space chart.
Covariant singularities can thus be studied via the field-space invariants. 
\par
With the purpose of being as general as possible, we shall leave the classical action unspecified. 
The metric $\bm G$ of the field space is supposed to be seen as part of the definition of the theory.
The line element in $\mathcal S$ is naturally defined by
\beq
\d s^2
=
G_{ab}\,\d\varphi^a\,\d\varphi^b
=
\int\d^4x
\int\d^4x'\,
G_{AB}(x,x')\,\d\varphi^A(x)\,\d\varphi^B(x')
\ .
\eeq
For pure gravity theories, $\mathcal S=\mathcal G$ and one identifies $\varphi^a = g^{\mu\nu}(x)$.
In this case, assuming the field-space metric to be ultralocal ({\em i.e.}~proportional
to the spacetime Dirac delta and independent of derivatives of the spacetime metric),
there is a unique {(up to a global factor)} one-parameter familiy of field-space metrics
\begin{equation}
G_{ab} = G_{AB} \, \delta(x,x')
\ ,
\label{eq:dw}
\end{equation}
where
\beq
G_{AB}
=
\frac12 
\left( g_{\mu\rho}\, g_{\sigma\nu}
+ g_{\mu\sigma}\, g_{\rho\nu}  
+ c \, g_{\mu\nu} \, g_{\rho\sigma}
\right)
\label{eq:dw2}
\eeq
is called the DeWitt field-space metric~\cite{DeWitt:1967yk,DeWitt:1967ub} and
involves only a dimensionless parameter $c$.
Its inverse is found by solving $G_{AB}\, G^{BC} = \delta^C_A$, which gives
\beq
G^{AB}
=
\frac12 
\left( g^{\mu\rho} \,g^{\sigma\nu}
+ g^{\mu\sigma}\, g^{\rho\nu} 
- \frac{2\, c}{2 + n \, c} \, g^{\mu\nu} \,g^{\rho\sigma} \right)
\ ,
\eeq
where $n$ is the dimension of the spacetime $\mathcal M$.
The DeWitt functional metric is thus invertible only for $c\neq -2/n$.
The parameter $c$ cannot be determined without some additional assumption.
For example, Vilkovisky~\cite{Vilkoviskii:1984un,Vilkovisky:1984st} suggested that $\bm G$
should be identified from the highest derivative term in the classical action.
In that case, $c=-1$ for the Einstein-Hilbert action, but it would be different for higher-derivative gravity.
We shall leave $c$ unspecified.
\par
The connection on $\mathcal S$ is of the Levi-Civita type:
\beq
\Gamma^a_{\ bc}
=
\Gamma^A_{\ BC} \, \delta(x_A,x_B) \,\delta(x_A,x_C)
\ ,
\eeq
with
\beq
\Gamma^A_{\ BC}
=
\frac12 \,G^{AD} 
\left(\partial_B G_{DC} + \partial_C G_{BD} - \partial_D G_{BC}\right)
\ .
\eeq
In particular, the Levi-Civita connection for the DeWitt functional metric~\eqref{eq:dw} reads
\begin{align}
\Gamma^A_{\ BC}
=
&
-{ \delta^{(\lambda}}_{(\mu} \, g_{\nu)(\rho} \, { \delta^{\tau)} }_{\sigma)}
+
\frac14\, g_{\mu\nu}\, \delta^{(\lambda}_\rho\, \delta^{\tau)}_\sigma
+
\frac14\, g_{\rho\sigma}\, \delta^{(\lambda}_\mu\, \delta^{\tau)}_\nu
\nonumber
\\
&
-
\frac{1}{2\,(2 + n\, c)}\, g^{\lambda\tau}\, g_{\mu ( \rho}\, g_{\sigma ) \nu}
-
\frac{c}{4\,(2 + n\, c)}\, g^{\lambda\tau}\, g_{\mu\nu}\, g_{\rho\sigma}
\ .
\end{align}
\par
The functional Riemann curvature is then defined in the usual way as
\beq
\mathcal R^a_{\ bcd}
=
\partial_c \Gamma^a_{\ db} - \partial_d \Gamma^a_{\ cb}
+ \Gamma^a_{\ ce}\,\Gamma^e_{\ db}
+ \Gamma^a_{\ de}\,\Gamma^e_{\ cb}
\ ,
\eeq
with $\mathcal R_{bd} = \mathcal R^a_{\ bad}$ and $\mathcal R = \mathcal R^a_{\ a}$
being the the functional Ricci tensor and functional Ricci scalar, respectively.
For the DeWitt functional metric, the Ricci tensor is given by 
\beq
\mathcal R_{AB}
=
\frac14 
\left(g_{\mu\nu}\,g_{\rho\sigma} - n \, g_{\mu(\rho}\,g_{\sigma)\nu} \right)
\label{eq:dwricci}
\eeq
and the Ricci scalar by
\beq
\mathcal R 
= \frac{n}{4} - \frac{n^2}{8} - \frac{n^3}{8}
\ .
\label{eq:dwriccis}
\eeq
We remark that both quantities are local in $\mathcal M$ because of Eq.~\eqref{eq:dw},
and can therefore be used to inspect the field space $\mathcal S$, like spacetime scalars
are used to probe $\mathcal M$.
\par
The standard practice in General Relativity to decide whether a singularity is removable or just a coordinate
singularity is to seek singularities in the curvature invariants, such as the Kretschmann scalar.
Since diffeomorphism invariants are the same in all coordinate systems, only
``true''~\footnote{As we advocate in this paper, true singularities must be invariant not only
under diffeomorphisms but also under field redefinitions.
That is the reason we wrote true between quotes.}
singularities, if any, would be manifested.
Analogously, we must seek a scalar functional defined in the field space in order to investigate
the appearance of singularities.
We shall define the functional Kretschmann scalar of the underlying field space $\mathcal S$ as
\beq
\mathcal K = \mathcal R_{ABCD}\,\mathcal R^{ABCD}
\ .
\eeq
The functional Kretschmann scalar is naturally invariant under field redefinitions,
it thus exhibits singularities that can neither be removed by field redefinitions nor spacetime
coordinate transformations.
To assess whether a singularity is real or only a consequence of a bad choice of field variables,
one must therefore calculate $\mathcal K$.
Note that the functional Kretschmann scalar depends only on the field-space metric $\bm G$.
The dependence on a particular action would become manifest once the components $G_{AB}$ 
are identified through the highest derivative term of the classical action as explained before.
For the DeWitt functional metric~\eqref{eq:dw2}, the dependence on the theory is thus encoded
by the parameter $c$.
\par
The difficulty now is to compute $\mathcal K$ explicitly.
For a spacetime $\mathcal M$ of dimension $n$, each capital index corresponds to
$(n^2 + n)/2$ degrees of freedom.
The functional Riemann tensor alone thus contains $(n^2 + n)^4/16$ components.
Futhermore, one must contract it with itself to obtain $\mathcal K$.
The use of a good computer algebra system is clearly very convenient
and we used the software {\tt Cadabra2}~\cite{Peeters:2018dyg,Peeters:2007wn}
to obtain the fairly simple result
\beq
\mathcal K
=
\frac{n}{8}\left(\frac{n^3}{4} + \frac{3 n^2}{4} - 1\right)
.
\label{KDW}
\eeq
This clearly shows that $\mathcal K$ is smooth for any spacetime metric $\bm g$
in any spacetime dimension $n$.
More importantly, $\mathcal K$ turns out to not depend on the DeWitt parameter $c$.
Therefore, \textit{every theory of pure gravity} is absent of curvature singularities in the
field space $\mathcal G$.
Note that, as we explained before, curvature invariants generally
depend on the spacetime metric, thus spacetime singularities are indeed related
to field-space singularities.
Because field-space curvature scalars are invariant under field redefinitions, any
singularity manifested in $\mathcal K$ would represent
singularities that cannot be removed by field redefinitions.
This is analogous to the fact that singularities in the spacetime Kretschmann
$K=R_{\mu\nu\rho\sigma}\,R^{\mu\nu\rho\sigma}$ cannot be removed by coordinate
transformations.
\par
{
As an example, let us make a slight modification of the DeWitt metric~\eqref{eq:dw2}
by including a global factor $g = \det(\bm g)$ for the determinant of the spacetime metric,
that is
\begin{equation}
    G_{A B}
    =
    \frac12 \left(-g\right)^\epsilon
    \left( g_{\mu\rho}\, g_{\sigma\nu}
        + g_{\mu\sigma}\, g_{\rho\nu}  
        + c \, g_{\mu\nu} \, g_{\rho\sigma}
    \right)
    \ ,
\end{equation} 
where $\epsilon$ is an arbitrary exponent.
In this case, the Kretschmann scalar becomes
\beq
    \mathcal K
    =
    \frac{n}{8}
    \left(-g\right)^{-2 \epsilon}
    \left(\frac{n^3}{4} + \frac{3 n^2}{4} - 1\right)
    \ .
\label{ex}
\eeq
Such a result signals the potential existence of field-space singularities at
$g = 0$ for $\epsilon > 0$ or at $g \to \infty$ for $\epsilon < 0$.
It is only for $\epsilon = 0$ that one obtains an everywhere-finite result,
in agreement with Eqs.~\eqref{eq:dw2} and \eqref{KDW}.
These potential singularities will, of course, only become actual singularities if,
given a theory (be it General Relativity or else), there exist solutions
of the corresponding equations of motion which satisfy $g = 0$ for the field-space
metric with $\epsilon > 0$ or $g \to \infty$ for $\epsilon < 0$.
Spacetime singularities for which $g$ is either finite (for $\epsilon < 0$)
or non-zero (for $\epsilon > 0$) can in principle be removed by field redefinitions.
A simple example that can be used to understand the dependence of the field-space
metric (here parameterized by $\epsilon$) on the presence of covariant singularities
is the Schwarzschild black hole,
\begin{equation}
    \d s^2
    =
    - \left( 1 - \frac{2MG}{r} \right) \d t^2
    + \left( 1 - \frac{2MG}{r} \right)^{-1} \d r^2
    + r^2 \left(\d\theta^2 + \sin\theta \,\d\phi^2\right)
    \ ,
    \label{eq:BH}
\end{equation}
whose determinant reads
\begin{equation}
    g = - r^4 \sin^2\theta
    \ .
    \label{detg}
\end{equation}
Assuming that the dynamics is described by General Relativity,
of which Eq.~\eqref{eq:BH} is a well-known solution, one can assess the presence
of covariant singularities by combining Eqs.~\eqref{ex} and \eqref{detg}.
Thus, if we take $\epsilon > 0$, a covariant singularity exists at the origin $r=0$,
whereas a covariant singularity is present at $r \to \infty$ for the $\epsilon < 0$.
No covariant singularity seems however to exist in the Schwarzschild black hole for $\epsilon = 0$.
Indeed, the fact that the Kretschmann scalar $\mathcal K$ is constant and finite
for $\epsilon=0$ suggests that no covariant singularity would be present,
regardless of the underlying theory, in this case.
Therefore, the absence of covariant singularities largely depends on the choice of the field-space metric.}
\par
It so happens that the Kretschmann scalar is everywhere finite for the DeWitt metric with $\epsilon=0$,
which suggests that all spacetime singularities could be removed by field redefinitions
in this case.
This means that, in pure gravity, spacetimes $\mathcal M_1$ with curvature singularities
signalled by the diverging Kretschmann scalar can always be extended to regular spacetimes
$\mathcal M_2$ which do not contain such singularities by a redefinition of the metric $\bm g$.
{
It thus seems that a field-space metric with $\epsilon=0$ is the optimal choice to avoid covariant singularities in pure gravity.}
\par
Note that the above discussion is restricted to curvature singularities.
Other types of singularities in field space, such as the conical ones, could very well be present
and be reflected in the physical observables.
The study of general singularities is left for future developments ({\em e.g,}~see Ref.~\cite{Casadio:2021rwj}).
\section{Conclusions}
\setcounter{equation}{0}
\label{Sconc}
In this paper, we have proposed to investigate singularities in the field space rather than in spacetime.
 This is particularly important in order to perform an analysis that is covariant under
field redefinitions.
Such a generalised form of the principle of covariance is indeed required in the study of quantum
effects in cosmology and astrophysics, where observables are in-in correlation functions rather
than $S$-matrix elements.
Although the equivalence theorem guarantees that scattering amplitudes are invariant under
the set of field redefinitions that keep the asymptotic states fixed, the same cannot be said
about in-in correlations functions, whose calculation requires the geometrical approach introduced by
Vilkovisky and DeWitt~\cite{Vilkovisky:1984st,DeWitt:1988dq}.
As a result of Vilkovisky and DeWitt's works, one obtains a formalism where fields are mere
coordinates in the field space and observables are thus scalars defined on the same space.
Spacetime singularities are field-dependent, thus their real significance is not clear until one
proves their presence for every choice of field variables.
Existing examples in the literature indeed show that certain singularities in spacetime can be removed
by field redefinitions albeit being non-removable under change of coordinates.
We corroborated such examples by remarking that the Hawking-Penrose theorem is not a covariant
result under field redefinitions.
\par
Finding field redefinitions that can eliminate singularities is obviously not always feasible in practice
as there are infinitely many choices of parameterisation for a field theory.
The most promising approach is to calculate curvature invariants in field space.
We showed that the Kretschmann scalar of the DeWitt functional metric turns out to be free of singularities
for {$\epsilon=0$}.
Surprisingly, this result does not depend on the choice of the gravitational action, which is encoded
in the parameter $c$ in Eq.~\eqref{eq:dw2}.
Any pure gravity theory is therefore devoid of curvature singularities {should one take $\epsilon=0$}.
\par
We should note that removing a singularity from a field configuration certainly makes
its description more complete, but the fact that a singularity is removable in field space does not
imply that there is no interesting physics occurring around it.
Horizons as removable spacetime singularities clearly teach us that investigating the physics
likely requires a case by case study.
In fact, studying specific models of self-gravitating systems is one of the natural developments
of the present work, as it is the formal analysis of field-space invariants for more general theories
than pure gravity.
We will pursue both directions in future works.
\section*{Acknowledgments}
This work is partially supported by the INFN grant FLAG.
The work of R.C.~has also been carried out in the framework of activities of the National Group of Mathematical Physics
(GNFM, INdAM) and COST action Cantata.
A.K.~was partially supported by the Russian Foundation for Basic Research grant No.~20-02-00411.
%
%
%
%

%
\end{document}